\documentclass[12pt]{article}
\usepackage{wrapfig}
\usepackage[pdftex]{graphicx}
\usepackage{float}
\usepackage[stable]{footmisc}
\usepackage{booktabs}
\usepackage{amsmath}
\usepackage{amssymb}
\usepackage{graphicx}
\usepackage{epsf}
\usepackage{fancyhdr}
\newcommand\pubnumber{}
\newcommand\pubdate{\today}

\def\napoli{$^{\rm A}$University of Tsukuba\\
$^{\rm B}$High Energy Accelerator Research Organization (KEK)\\ 
$^{\rm C}$ Institute of Physical and Chemical Research (REKEN) \\}

\def\Title#1{\begin{center} {\Large #1 } \end{center}}
\def\Author#1{\begin{center}{ \sc #1} \end{center}}
\def\Address#1{\begin{center}{ \it #1} \end{center}}

\newcommand\pubblock{\rightline{\begin{tabular}{l} \pubnumber\\
\pubdate \end{tabular}}}
\newenvironment{Abstract}{\begin{quotation} }{\end{quotation}}
\newenvironment{Presented}{\begin{quotation} \begin{center} 
PRESENTED AT\end{center}\bigskip 
\begin{center}\begin{large}}{\end{large}\end{center} \end{quotation}}
\def\Acknowledgements{\bigskip \bigskip \begin{center} \begin{large}
\bf ACKNOWLEDGEMENTS \end{large}\end{center}}
\def\beq{\begin{equation}}
\def\eeq#1{\label{#1}\end{equation}}
\def\eeqn{\end{equation}}

\def\beqa{\begin{eqnarray}}
\def\eeqa#1{\label{#1}\end{eqnarray}}
\def\eeqan{\end{eqnarray}}

\let\bar=\overbar

\def\Dslash{\not{\hbox{\kern-4pt $D$}}}
\def\dslash{\not{\hbox{\kern-2pt $\del$}}}

\def\msb{{\bar{\ssstyle M \kern -1pt S}}}

\begin{document}
\begin{titlepage}
\pubblock

\vfill
\Title{Compensation for TID Damage in SOI Pixel Devices}
\vfill
\Author{Naoshi Tobita$^{\rm A}$, Shunsuke Honda$^{\rm A}$, Kazuhiko Hara$^{\rm A}$, Wataru Aoyagi $^{\rm A}$, Yasuo Arai$^{\rm B}$, Toshinobu Miyoshi$^{\rm B}$, Ikuo Kurachi$^{\rm B}$, Takaki Hatsui$^{\rm C}$, Togo Kudo$^{\rm C}$, Kazuo Kobayashi$^{\rm C}$ }
\Address{\napoli}
\vfill

\begin{Abstract}
We are investigating adaption of SOI pixel devices for future high energy physic(HEP) experiments.
The pixel sensors are required to be operational in very severe radiation environment. 
Most challenging issue in the adoption is the TID (total ionizing dose) damage 
where holes trapped in oxide layers affect the operation of nearby transistors. 
We have introduced a second SOI layer - SOI2 beneath the BOX (Buried OXide) layer - 
in order to compensate for the TID effect by applying a negative voltage to this  electrode to
cancel the effect caused by accumulated positive holes. 
In this paper, the TID effects caused by $\rm {^{60}Co}$ ${\rm \gamma}$-ray irradiation are presented 
based on the transistor characteristics measurements. 
The irradiation was carried out in various biasing conditions to investigate hole accumulation dependence 
on the potential configurations.
We also compare the data with samples irradiated with X-ray. 
Since we observed a fair agreement between the two irradiation datasets, the TID effects have been investigated in a wide dose range from 100~Gy to 2~MGy.  
\end{Abstract}

\vfill
\begin{Presented}
International Workshop on SOI Pixel Detector (SOIPIX2015), Tohoku University, Sendai, Japan, 3-6, June, 2015.
\end{Presented}
\vfill
\end{titlepage}
\def\thefootnote{\fnsymbol{footnote}}
\setcounter{footnote}{0}

%%%----------------------------------------------------------------------------

 \section{Introduction}

The Silicon-On-Insulator (SOI) pixel detectors are being developed for various applications \cite{r:SOI}. 
Applications in high-energy physics experiments require a couple of issues to be cleared. 
In the SOI devices the total-ionization-dose (TID) effect is substantial in severe radiation environments, 
where the trapped holes in the insulator affect the operation of the transistors located close by. 
 The annual doses at the inner-most layer in typical HEP experiments are; 158~kGy/y at LHC-ATLAS \cite{r:ATLAS}, 1.6~MGy/y at HL-LHC, and 1~kGy/y at ILC-ILD \cite{r:ILC}.

We are studying TID compensation by using double SOI, where the potential of the middle silicon layer we have introduced additionally is controlled to cancel the effect, e.g., applying negative voltage to the SOI2 (${\rm V_{SOI2}}$) cancels the effects caused by accumulated positive holes \cite{r:Hara}\cite{r:Honda}.
The hole accumulation, however, should be dependent on the configuration of the terminal biases applied during irradiation. The purpose of this study is to investigate the compensation dependence on such bias configurations.
A general study concerning the influence of biasing during irradiation can be found in Tab~.\ref{Biasingcondition}.
%
%\begin{figure}[htbp]
%\vspace{0.01in}
%\begin{center}
%\begin{tabular}{c}
%\includegraphics[width=80mm]{./fig/DoubleSOI.png}
%\end{tabular}
%\caption{Middle SOI layer, called SOI2, have been introduced to compensate the TID damage by applying negative voltage}
%\label{DoubleSOI}
%\end{center}
%\end{figure}

  %  %  %%%----------------------------------------------------------------------------

\section{Samples}
The TID compensation was evaluated by irradiating TrTEG (Transistor Element Group) samples up to 2 MGy with $\rm {^{60}Co}$ ${\rm \gamma}$'s. 
Various types of transistors were fabricated in TrTEG chips for this study. \\
TrTEG6 has 18 types transistors (L/W values, two gate oxide thicknesses core/IO, body connection schema) for each of NMOS and PMOS,as summarized in Tab.~\ref{TrTEG6tr}. 
In this note we concentrate on their L/W dependence.
TrTEG7 has seven types for each of NMOS and PMOS which were irradiated under various biasing conditions. 

\begin{figure}[htbp]
\vspace{0.01in}
\begin{center}
\begin{tabular}{ll}
\includegraphics[width=55mm]{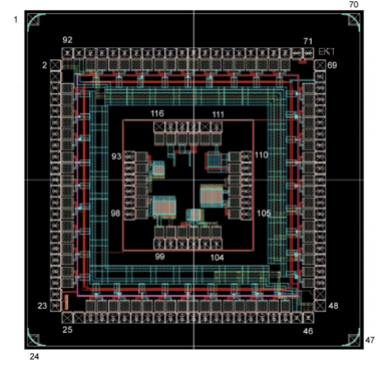}
\includegraphics[width=55mm]{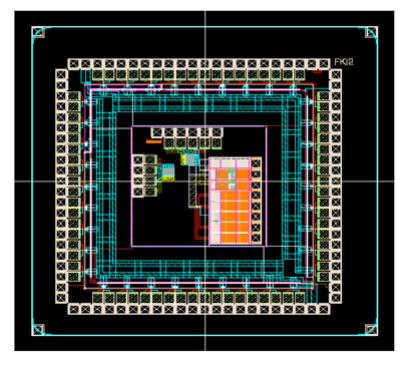}
\end{tabular}
\caption{(left) TrTEG6 and (right) TrTEG7. TrTEG6 has 18 types of transistors for each NMOS and PMOS. 
TrTEG7 samples were irradiated in seven biasing conditions.}
\label{TrTEG}
\end{center}
\end{figure}

Fig.\ref{TrTEG} illustrates drawings of the TrTEG6 and TrTEG7 chips.
The chips were mounted on ceramic chip carriers and wire-bonded for biasing.
They were irradiated with $\rm {^{60}Co}$ at JAEA (Takasaki) \cite{r:ATOM}.
The total  dose ranged from 3 kGy up to 2 MGy for TrTEG6 samples but with all electrode grounded.
The TrTEG7 samples were irradiated to 100 kGy but with various biasing conditions, see Tab.~\ref{Biasingcondition}.
In total seven types of biasing conditions were examined. 
This paper covers four types. 

The dose rate was from 0.2~kGy/h up to 10~kGy/h. 
All samples were kept at room temperature during the irradiation. 
The samples were brought to University of Tsukuba in four hours in room temperature, 
then kept at -20Ž in the refrigerator except during measurement.

\begin{table}[H]
\caption{TrTEG6 transistors parameters}
\label{TrTEG6tr}
\begin{center}
\setlength{\tabcolsep}{5pt}
\footnotesize
\begin{tabular}{|c|c|c|c|c|c|} \hline
Tr & L [${\mu m}$] & W [${\mu m}$] & Gate type &  ${\rm V_{th}}$ & Body-Connection \\ \hline 
0 & 0.2 & 5 & Core & Normal & Floating \\
1 & 0.5 & 5 & Core & Normal & Floating \\
2 & 1.0 & 5 & Core & Normal & Floating \\
3 & 0.2 & 5 & Core & Low & Floating \\
4 & 0.5 & 5 & Core & Low & Floating \\
5 & 1.0 & 5 & Core & Low & Floating \\
6 & 0.35 & 5 & IO & high & Floating \\
7 & 0.35 & 5 & IO & Low & Floating \\
8 & 0.2 & 5 & Core & Low & Source-Tie \\
9 & 0.5 & 5 & Core & Low & Source-Tie \\
10 & 1.0 & 5 & Core & Low & Source-Tie \\
11 & 0.4 & 10 & Core & Normal & Source-Tie2 \\
12 & 0.6 & 6 & Core & Normal & Source-Tie2 \\
13 & 1.0 & 10 & Core & Normal & Source-Tie2 \\
14 & 0.2 & 5 & Core & Normal & Body-Tie \\
15 & 0.5 & 5 & Core & Normal & Body-Tie \\
16 & 1.0 & 5 & Core & Normal & Body-Tie \\
17 & 1.0 & 5 & IO & Normal & Source-Tie \\ \hline

\end{tabular}
\end{center}
\end{table}

\begin{table}[H]
\caption{Biasing conditions during TrTEG7 irradiation}
\label{Biasingcondition}
\begin{center}
\setlength{\tabcolsep}{5pt}
\footnotesize
\begin{tabular}{|c|c|c|c|c|} \hline
\phantom & ${\rm V_{s}}$ [V] &  ${\rm V_{d}}$ [V]  &  ${\rm V_{g}}$ [V]  &  ${\rm V_{SOI2}}$ [V] \\ \hline 
Sample1 & 0 & 0 & 0 & 0 \\
Sample2 & 0 & 0 & 0 & -5 \\
Sample3 & 0(1.8) & 1.8(0) & 0(1.8) & -5 \\
Sample4 & 0(1.8) & 1.8(0) & 1.8(0) & -5 \\ \hline 
\end{tabular}
\end{center}
\end{table}

\section{{\rm $I_d-V_g$} Curve}
\begin{figure}[htbp!]
\vspace{0.01in}
\begin{center}
\begin{tabular}{c}
\includegraphics[width=140mm]{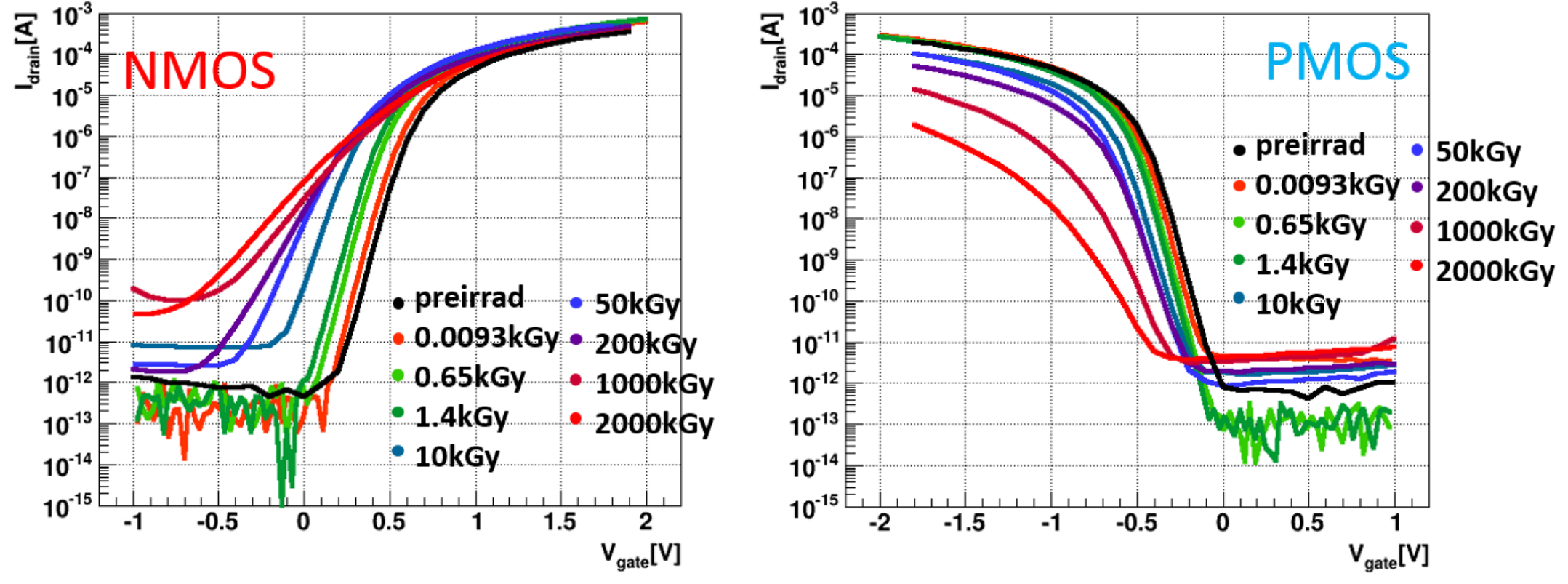}
\end{tabular}
\caption{${\rm I_{d}}$-${\rm V_{g}}$ curves for (left) NMOS and (right) PMOS transistors with L$=1.0~{\mu}$m, 
W$=5.0~{\mu}$m, Core, Low ${\rm V_{th}}$, Source-Tie). 
They were irradiated with ${\rm V_{SOI2}}=0$. 
Both NMOS and PMOS I-V curves shift negatively as accumulating the dose gradually from pre-irradiation curve (black) to 2~MGy curve (red).}
\label{IVCurve1}
\end{center}
\end{figure}

We measured {\rm $I_d-V_g$} curve to evaluate characteristics of the transistors.
The obtained {\rm $I_d-V_g$} curves are shown in Fig.\ref{IVCurve1}
as a function of dose for typical NMOS and PMOS transistors. 
The {\rm $I_d-V_g$} curve was measured at ${\rm V_{d} = 1.8~V}$, ${\rm V_{s} = 0~V}$ for NMOS, and ${\rm V_{d} = 0~V}$, ${\rm V_{s} = 1.8~V}$ for PMOS with ${\rm V_{BPW}}$ grounded, ${\rm V_{BACK}}$ floating, and ${\rm V_{g} = \rm V_{s}}$ for body-tie type transistors.
As shown the figure, the {\rm $I_d-V_g$} curve shifts negatively as accumulating the dose. 
Pre-irradiation curve as shown in black as a reference.

\begin{figure}[htbp!]
\vspace{0.2in}
\begin{center}
\begin{tabular}{c}
\includegraphics[width=140mm]{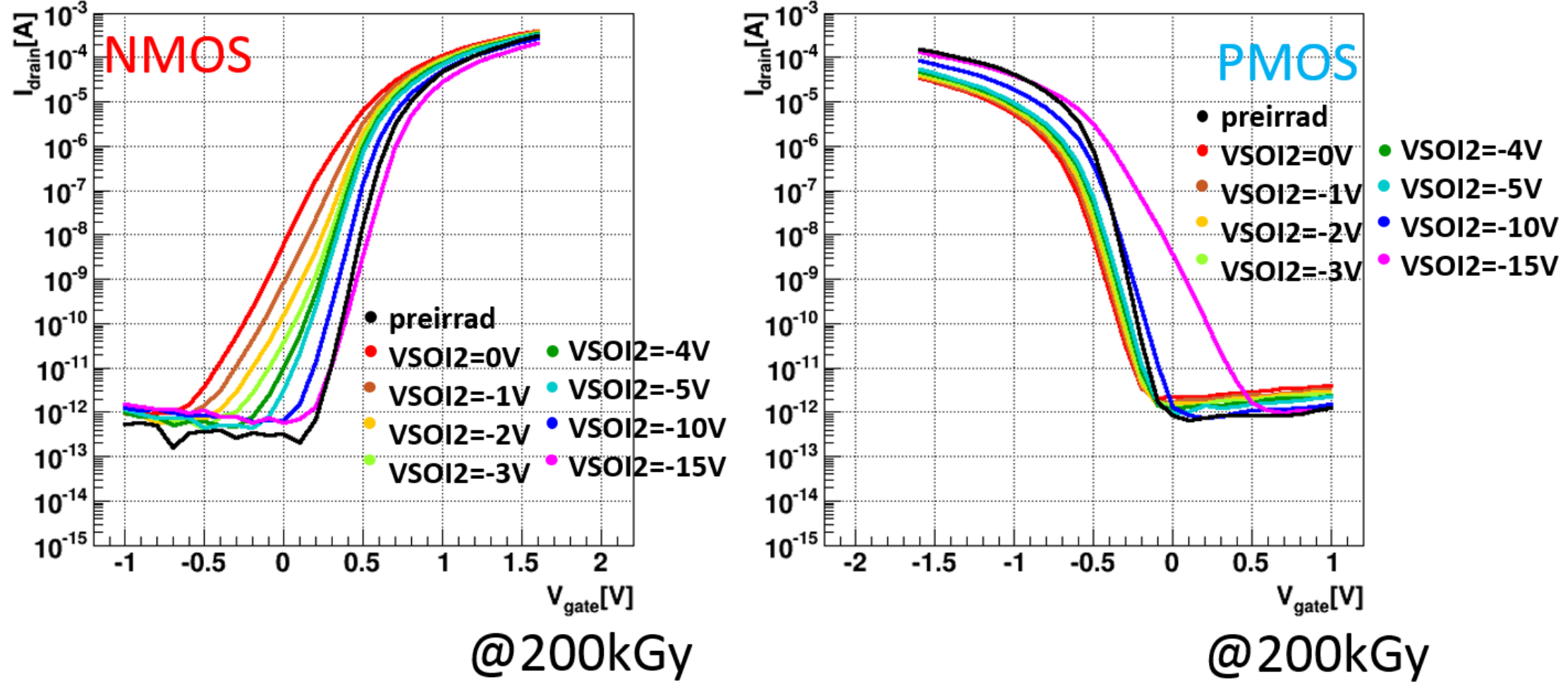}
\end{tabular}
\caption{${\rm I_{d}}$-${\rm V_{g}}$ curves for (left) NMOS and (right) PMOS measured with changing ${\rm V_{SOI2}}$ voltage from 0~V to -15~V. The samples were irradiated to 200~kGy.
The I-V curves shifted by irradiation (0~V, red) are compensated back to the pre-irradiation curves (black) by adjusting ${\rm V_{SOI2}}$ voltage (for example at -10~V, blue).}
\label{IVCurve2}
\end{center}
\end{figure}

Fig.~\ref{IVCurve2} shows the curves of samples irradiated to 200~kGy but with changing the SOI2 voltage(${\rm V_{SOI2}}$) from 0~V to $-$15~V. By adjusting ${\rm V_{SOI2}}$ to $-$10~V the irradiated I-V curve is nearly compensated back to the pre-irradiation curve.

\vspace{0.01in}

 \section{Threshold ${\rm V_{th}}$ and trans-conductance ${\rm g_{m}}$}

\begin{figure}[htbp]
\vspace{0.01in}
\begin{center}
\begin{tabular}{c}
\includegraphics[width=160mm]{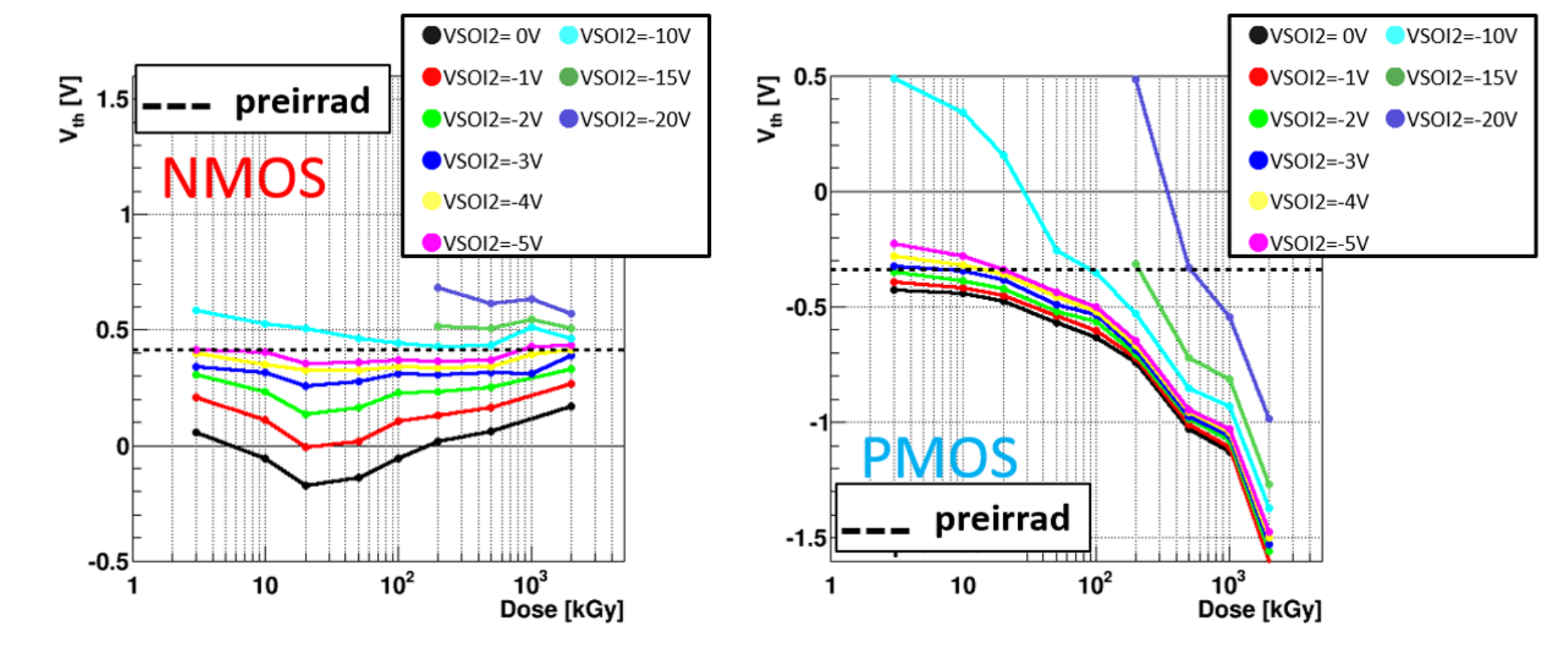}
\end{tabular}
\caption{The threshold voltage (${\rm V_{th}}$) shifts are shown for (left) NMOS and (right) PMOS. 
Without applying ${\rm V_{SOI2}}$, they shift negatively and substantially 
as ${\rm V_{SOI2}}$ = 0~V curve (black) shows. 
By applying ${\rm V_{SOI2}}$, the threshold voltage can be set back to the pre-irradiation value (dotted lined).}
\label{Vth}
\end{center}
\end{figure}

We define two values to evaluate transistor characteristics quantitatively, 
threshold voltage (${\rm V_{th}}$) and trans-conductance (${\rm g_{m}}$). 
The ${\rm V_{th}}$ is defined as ${\rm V_{g}}$ at ${\rm I_{d}}$ = 100[nA]~W/L (W: gate width, L: gate length). 
The evolution with the dose of the threshold voltage of a typical transistor is plotted 
in Fig.~\ref{Vth}.
The curves are shown for various ${\rm V_{SOI2}}$ settings applied during measurement (they were grounded during irradiation for these samples). 
The dotted line indicates the pre-irradiation threshold voltage. 
By adjusting ${\rm V_{SOI2}}$, the threshold voltage can be set back to the pre-irradiation value.

\begin{figure}[htbp]
\vspace{0.01in}
\begin{center}
\begin{tabular}{c}
\includegraphics[width=160mm]{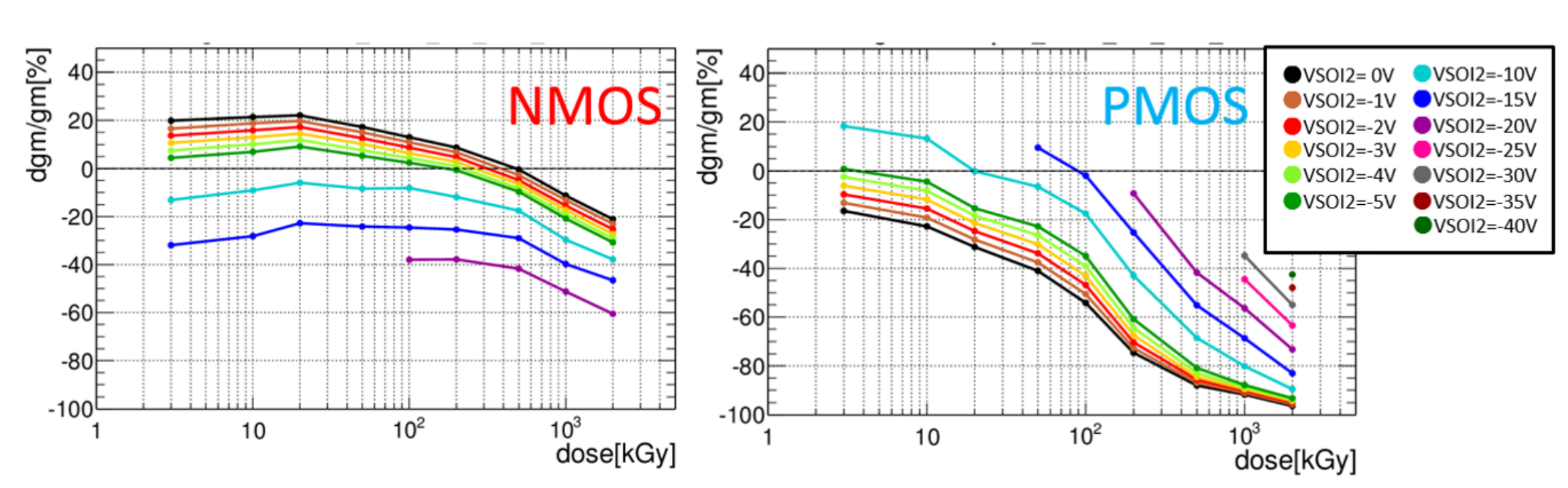}
\end{tabular}
\caption{The relative shift (d${\rm g_{m}}$/${\rm g_{m}}$) with dose for some ${\rm V_{SOI2}}$ voltages for(left)  NMOS and (right) PMOS. 
For NMOS, optimum ${\rm V_{SOI2}}$ can be found for ${\rm g_{m}}$ compensation below 500~kGy, 
while PMOS degradation is substantially and large ${\rm V_{SOI2}}$ voltage is required.
The compensation at large dose (100~kGy or higher) seems become difficult.}
\label{gm}
\end{center}
\end{figure}

The trans-conductance ${\rm g_{m}}$ is defined as ${\rm g_{m} = (\partial \rm I_{d}/\partial \rm V_{g})}$. The relative change  (d${\rm g_{m}}$/${\rm g_{m}}$) as defined in Equation (1)
are shown in Fig.\ref{gm} for some ${\rm V_{SOI2}}$ voltages. 
In general  NMOS can find an appropriate ${\rm V_{SOI2}}$ below 500~kGy, 
while PMOS requires relatively large ${\rm V_{SOI2}}$ for compensation.

In a recent study \cite{r:Kurachi}, we have verified that the large  ${\rm g_{m}}$ degradation in PMOS can be minimized substantially by adjusting the dose profile in gate fabrication. 
Further studies are in progress. 

%\[\frac{x}{x-1}\]
\begin{equation}
\frac{d\rm g_{m}}{\rm g_{m}}=\frac{\rm g_{m}(\rm V_{SOI2},dose)-\rm g_{m}(\rm V_{SOI2},preirrad)}{\rm g_{m}(\rm V_{SOI2},preirrad)}\times100
%(\rm g_{m} = \frac{\partial \rm I_{d}}{\partial \rm V_{g}})
\end{equation}

\section{L-dependence of ${\rm V_{th}}$ shift}

\begin{figure}[htbp]
\vspace{0.01in}
\begin{center}
\begin{tabular}{c}
\includegraphics[width=140mm]{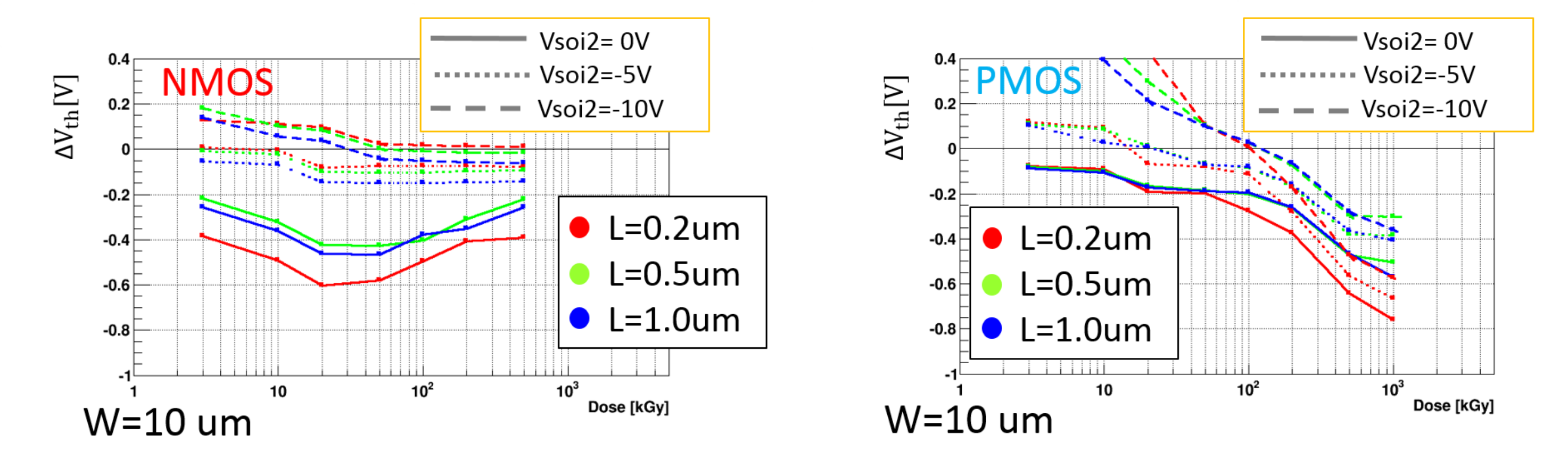}
\end{tabular}
\caption{The threshold shift-dependence on the transistor gate length L for (left) NMOS and (right) PMOS. 
Shorter length (L = 0.2${\mu}$m, solid red) transistors are more sensitive to TID effect. 
However, by applying ${\rm V_{SOI2}}$ for compensation (dotted lines with color), 
the difference of ${\rm V_{th}}$ shift among three L values becomes small.}
\label{L-length}
\end{center}
\end{figure}

We have evaluated the TID effect dependence on the transistor length  L to investigate possible edge effect enhancement in shorter devices. 
We define ${\Delta \rm V_{th}}$ as ${\Delta \rm V_{th}}$ = (${\rm V_{th-dose}}$-${\rm V_{th-preirrad}}$)/${\rm V_{th-preirrad}}$ for transistors with various L. 
As shown in Fig.~\ref{L-length}, shorter the L, larger the threshold voltage shift caused by TID for both NMOS and PMOS. 
The L dependence, however, diminishes by applying ${\rm V_{SOI2}}$ for compensation - the difference among three L values 0.2, 0.5, 1.0~$\mu$m is small when compensated.  

\section{${\rm V_{th}}$ shift per biasing condition during irradiation}

\begin{figure}[htbp]
\vspace{0.01in}
\begin{center}
\begin{tabular}{c}
\includegraphics[width=140mm]{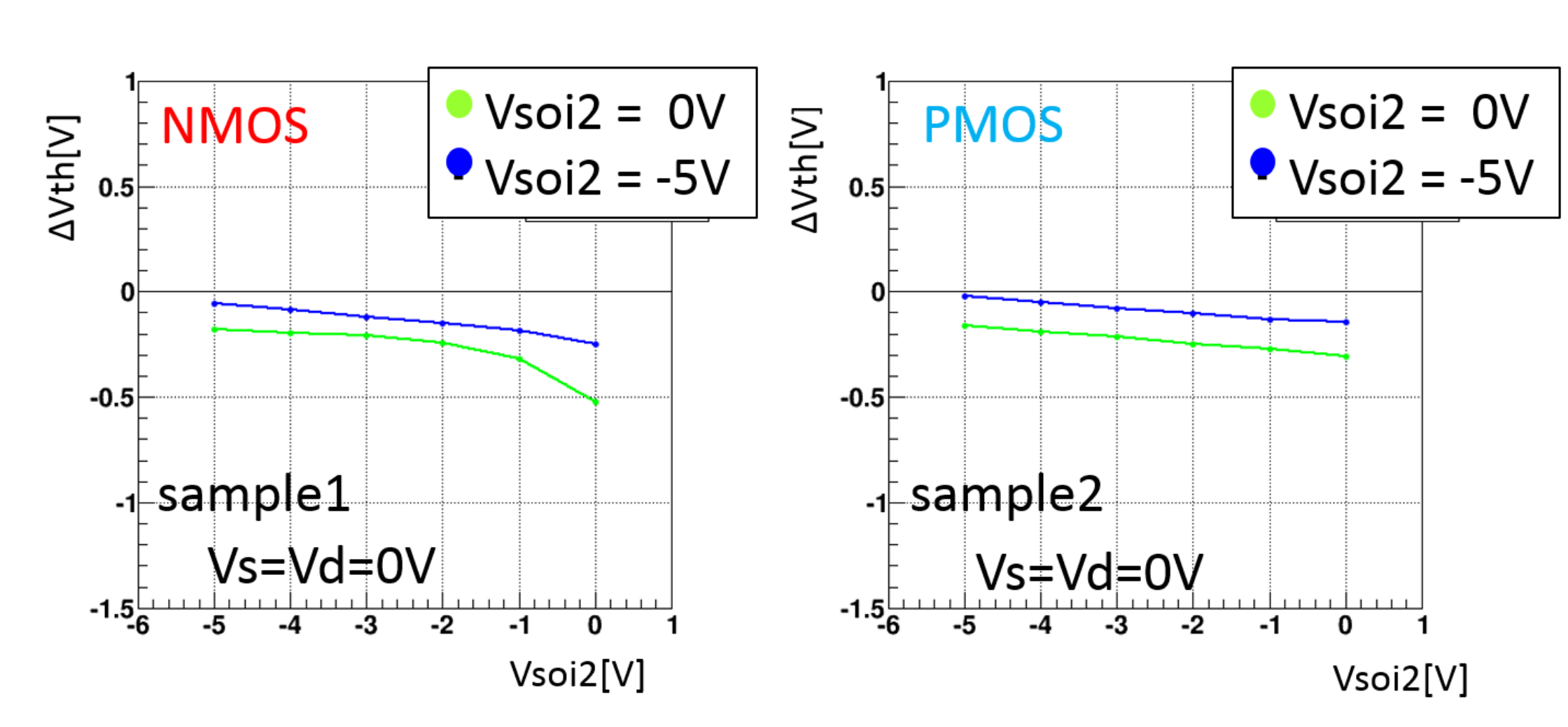}
\end{tabular}
\caption{The threshold voltage shifts measured with changing ${\rm V_{SOI2}}$ for transistors 
(L$=0.6~{\mu}$m, W$=60~{\mu}$m, Core, N${\rm V_{th}}$, ST2) of (left) NMOS and (right) PMOS.
 All terminals grounded except ${\rm V_{SOI2}}$ during irradiation. 
The shifts for the samples with ${\rm V_{SOI2}}$ -5~V (blue) during irradiation 
are reduced compared to the grounded sample (green).}
\label{bias-dependence1}
\end{center}
\end{figure}

\begin{figure}[htbp]
\vspace{0.01in}
\begin{center}
\begin{tabular}{c}
\includegraphics[width=140mm]{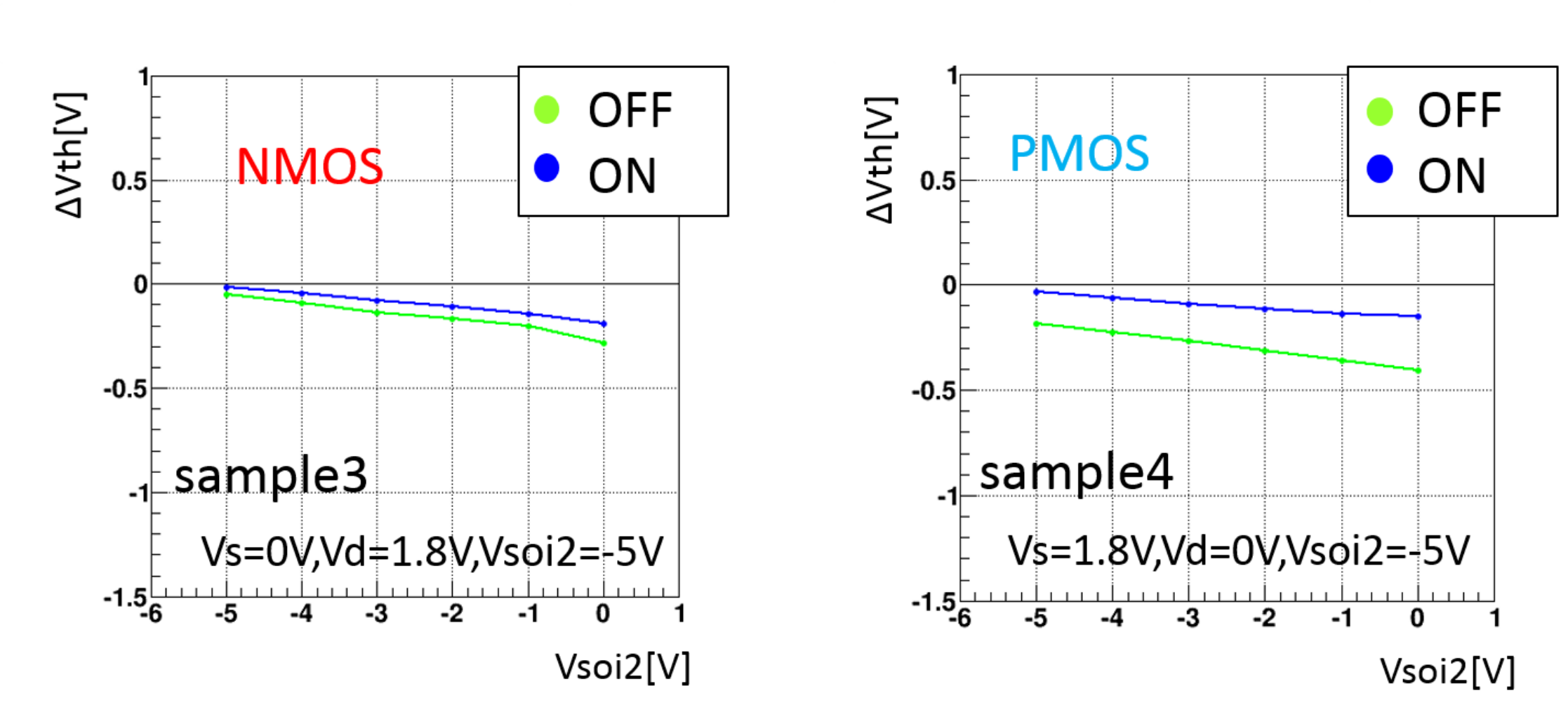}
\end{tabular}
\caption{The threshold voltage shifts measured with changing ${\rm V_{SOI2}}$. The transistor parameters are identical to Fig.~\ref{bias-dependence1}.  The transistors were either on of off during irradiation with applying ${\rm V_{SOI2}}$ of -5~V. 
The on-state (blue) transistors show smaller threshold voltage shift than the off-state (green) transistors.}
\label{bias-dependence2}
\end{center}
\end{figure}

The threshold voltage shifts measured with changing ${\rm V_{SOI2}}$(post-irradiation) are shown in Figs.~\ref{bias-dependence1} and \ref{bias-dependence2}. The data are compared between different biasing conditions  during irradiation. 
The plots are shown only for typical transistors, but all other transistors showed similar tendency.
In Fig.~\ref{bias-dependence1}, the samples were irradiated with all terminals grounded except that ${\rm V_{SOI2}}$ was either 0 V (green) or -5 V (blue) during irradiation. Fig.~\ref{bias-dependence2} is for the case the transistors were either in transistor-Off (green) or in transistor-On (blue) with ${\rm V_{SOI2}}$ was -5 V during irradiation.

Applying ${\rm V_{SOI2}}$ during irradiation reduces the threshold shift because hole traps in the oxide are less in the gate side attracted by negative ${\rm V_{SOI2}}$. 
The threshold shifts are smaller if the transistor is on state during irradiation. 
The effect is more significant for PMOS transistors.

\section{Comparison with X-ray irradiation}

\begin{figure}[here]
\vspace{0.01in}
\begin{minipage}{0.5\hsize}
\begin{center}
\includegraphics[width=50mm]{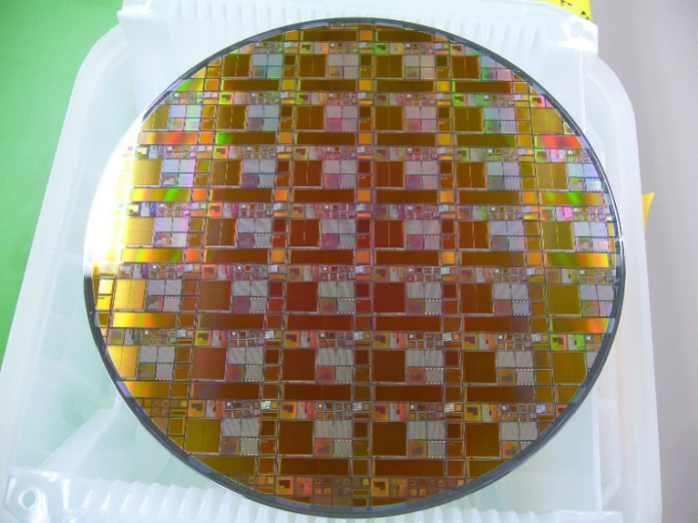}
\caption{Photo of a wafer of Radiation Test Element Group(RadTEG) \cite{r:RIKEN}.}
\label{RadTEG}
\end{center}
\end{minipage}
\hspace{0.2in}
\begin{minipage}{0.5\hsize}
\begin{center}
\includegraphics[width=70mm]{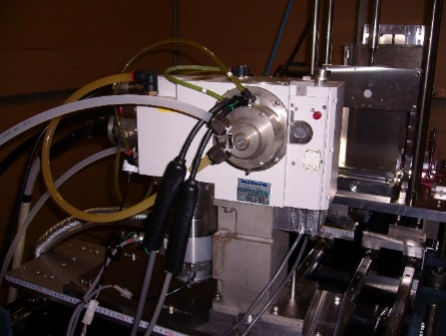}
\caption{X-ray irradiation using a Mo target with the tube current 100~mA and oltage 40~kV.\cite{r:RIKEN}}
\label{RadTEG2}
\end{center}
\end{minipage}
\end{figure}

While the above samples were irradiated with {\rm $^{60}$Co} covering a high dose range, more systematic irradiation has been performed with X-ray covering a dose as low as 100~Gy and up to 100~kGy.
The transistor parameters are summarized in Tab.~\ref{RadTEGtr}.
The irradiation was made on RadTEG samples \cite{r:RIKEN} fabricated on single SOI wafers. 
There are 8,000 transistors on a wafer.
The X-ray irradiation was carried out on a wafer basis, see Fig.~\ref{RadTEG}, with scanning the position unifrmly (see Fig.~\ref{RadTEG2}).

\begin{table}[H]
\caption{RadTEG transistor parameters\cite{r:RIKEN}}
\label{RadTEGtr}
\begin{center}
\setlength{\tabcolsep}{5pt}
\footnotesize
\tiny
\begin{tabular}{|c|c|c|c|c|c|} \hline
Gate type & Body-Connection & Max.L/W  & Min.L/W & ${\rm V_{th}}$ & ${\sharp}$ of Tr  \\ \hline 
Core & Floating & 0.2/0.5 & 10/10 & Low & 24 \\
Core & Floating & 0.2/0.5 & 10/10 & Normal & 24 \\
Core & Source-Tie & 0.2/0.2 & 5/5 & Low & 18 \\
Core & Source-Tie & 0.2/0.63 & 5/5 & Normal & 18 \\
Core & Source-Tie2 & 0.4/1.0 & 5/5 & Low & 18 \\
Core & Source-Tie2 & 0.4/1.0 & 5/5 & Normal & 18 \\
Core & Floating(enclosed) & 0.35/3.04 & 5/10 & Low & 12 \\
Core & Floating(enclosed) & 0.35/3.04 & 5/10 & Normal & 12 \\
Core & Body-Tie & 0.2/0.63 & 5/5 & Normal & 12 \\
Core & Body-Tie & 0.2/0.63 & 5/5 & Normal & 12 \\
IO & Floating & 0.35/0.5 & 10/10 & High & 24 \\
IO & Floating & 0.35/0.5 & 10/10 & Normal & 24 \\
IO & Source-Tie & 0.2/0.63 & 10/10 & Normal & 24 \\
IO & Source-Tie2 & 0.4/1.0 & 5/10 & Normal & 24 \\
IO & Floating(enclosed) & 0.35/3.04 & 10/10 & Normal & 24 \\
IO & Body-Tie & 0.2/0.63 & 5/5 & Normal & 12 \\ \hline
\end{tabular}
\end{center}
\end{table}

\begin{figure}[htbp]
\vspace{0.01in}
\begin{center}
\begin{tabular}{c}
\includegraphics[width=100mm]{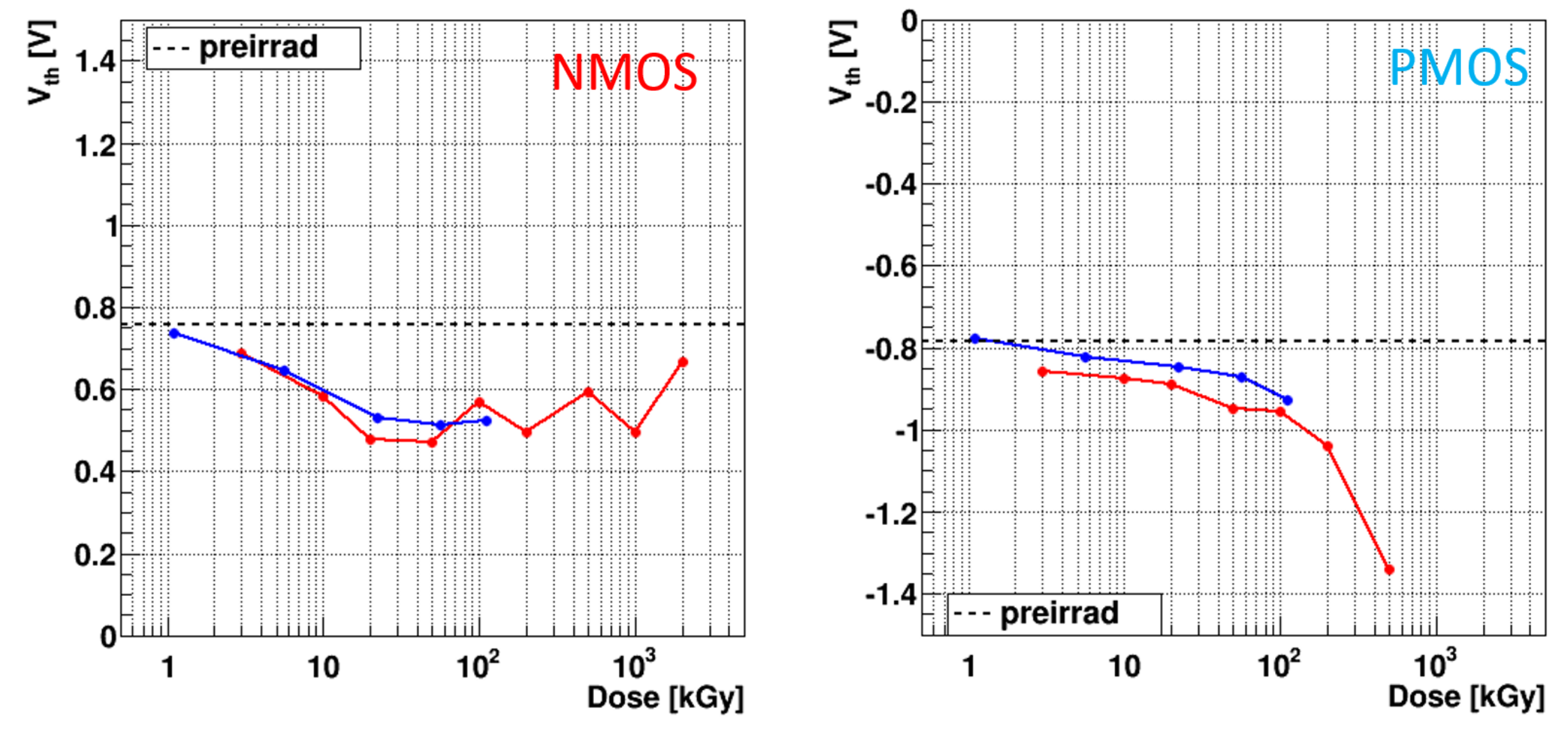}
\end{tabular}
\caption{${\rm V_{th}}$ shifts of (left) NMOS and (right) PMOS transistors in comparison of ${\rm \gamma}$-ray (red) and X-ray (blue) irradiations. }
\label{RadTEG-TrTEG}
\end{center}
\end{figure}

We compare the ${\rm V_{th}}$ shifts caused by two types of irradiation sources, see Fig.~\ref{RadTEG-TrTEG}. 
The transistors having identical parameters are chosen from the TrTEG6 and RadTEG samples.
The results from the $\rm {^{60}Co}$ irradiation are in fairly good agreement with the X-ray irradiation results in the overlapping dose region. Since RadTEG covers a lower dose range (down to 93Gy, not shown in this plot), the combined results provide us comprehensive understanding of the TID effects in wider dose range. Note that the BOX thicknesses are 200 nm for single SOI and about 160 nm for double SOI wafers, 
hence we expect some difference. In addition, the trapped hole fraction is larger with ${\rm{^{60}Co}}$ irradiaiton than with X-ray \cite{r:RIKEN2}, which may explain a general tendency that the threshold shifts are slightly larger in ${\rm {^{60}Co}}$ irradiated samples.  

\section{Summary}

We have investigated the TID effects in double SOI devices, evaluating changes in {\rm $I_d-V_g$} characteristics using TrTEGs.   By applying negative voltage to ${\rm V_{SOI2}}$, the transistor characteristics are compensated back to the pre-irradiation characteristics, in the dose range typically below 500~kGy.
 Detailed studies concerning the biasing conditions during irradiation have been presented. 
Such a dose range is below expected at the ILC pixel detector. The $\rm {^{60}Co}$ results are in fairly good agreement with X-ray irradiation, the both datasets covering the dose range from 100~Gy to 2~MGy.

\Acknowledgements

The authors are grateful for fruitful collaboration with the Lapis Semiconductor Co. Ltd. The double SOI wafers have been realized through their excellence. This work was supported by JSPS KAKENHI Grand Number 25109006, by KEK Detector
Technology Project and also by VLSI Design and Education Center (VDEC), The University of
Tokyo, with the collaboration of the Cadence Corporation and Mentor Graphics Corporation.

\end{document}